\documentclass[useAMS,usenatbib]{mn2e}
\usepackage{graphicx}

\title[Rapid oscillations in CVs -- V]{Dwarf nova oscillations and quasi-periodic oscillations in cataclysmic variables -- V. Results from an extensive survey}
\author[M.L. Pretorius, B. Warner and P.A. Woudt]{Magaretha L. Pretorius,$^{1}$\thanks{E-mail: mlp@astro.soton.ac.uk (MLP);~warner@physci.uct.ac.za (BW); pwoudt@circinus.ast.uct.ac.za (PAW)} Brian Warner\footnotemark[1] and Patrick A. Woudt\footnotemark[1]\\
Department of Astronomy, University of Cape Town, Private Bag, Rondebosch 7700, South Africa\\
$^{1}$Present address: School of Physics and Astronomy, University of Southampton, Southampton SO17 1BJ, United Kingdom}
\begin{document}


\pagerange{\pageref{firstpage}--\pageref{lastpage}} \pubyear{}

\maketitle

\label{firstpage}

\begin{abstract}
We present observations of dwarf nova oscillations (DNOs), longer period dwarf nova oscillations (lpDNOs), and quasi-periodic oscillations (QPOs) in 13 cataclysmic variable stars (CVs).  In the six systems WW Cet, BP CrA, BR Lup, HP Nor, AG Hya, and V1193 Ori, rapid, quasi-coherent oscillations are detected for the first time.  For the remainder of the systems discussed we have observed more classes of oscillations, in addition to the rapid oscillations they were already known to display, or previously unknown aspects of the behaviour of oscillations.  The period of a QPO in RU Peg is seen to change by 84\% over ten nights of the decline from outburst---the largest evolution of a QPO period observed to date.  A period-luminosity relation similar to the relation that has long been known to apply to DNOs is found for lpDNOs in X Leo; this is the first clear case of lpDNO frequency scaling with accretion luminosity.  WX Hyi and V893 Sco are added to the small list of dwarf novae that have shown oscillations in quiescence.
\end{abstract}


\section{Introduction}

Cataclysmic variable stars (CVs) vary in brightness on time scales from seconds to millions of years; the variability ranges in coherence from highly stable orbital and white dwarf spin cycles to aperiodic flickering.  A frequently observed property of CVs that has proved to be one of the most difficult to explain satisfactorily is the quasi-coherent variability seen on time scales of seconds to tens of minutes.  Dwarf nova oscillations (DNOs) and quasi-periodic oscillations (QPOs) were discovered a few decades ago (\citealt{p38}; \citealt{p11}).  More recently, Warner, Woudt \& Pretorius (2003, hereafter Paper III) introduced a third class---the longer period dwarf nova oscillations (lpDNOs).  The phenomenology of DNOs, lpDNOs, and QPOs, together with models to describe them, are reviewed by Warner (2004, hereafter W04).  Amongst the most important characteristics of DNOs are the period-luminosity relation (DNO frequency scales with accretion luminosity) and the time scales (typical DNO periods are $P_{DNO} \sim 10$~s).  The periods of normal QPOs ($P_{QPO}$) are longer, and the relation $P_{QPO}\approx 16\times P_{DNO}$ holds for DNOs and QPOs in a given system.  The period of an lpDNO ($P_{lpDNO}$) is typically midway between $P_{DNO}$ and $P_{QPO}$ in that system, i.e. $P_{lpDNO}\approx 4\times P_{DNO} \approx 1/4\times P_{QPO}$.  Rapid oscillations are almost exclusively associated with CVs in states of high mass transfer rate ($\dot{M}$), and are usually not observed in quiescent dwarf novae.  W04 also pointed out that some systems have QPOs at two separate time scales: QPOs with periods $\sim 1\,000$~s are seen in addition to QPOs that follow the relation $P_{QPO}\approx 16\times P_{DNO}$.   We will refer to the second group of QPOs as kilosecond QPOs.

Since rapid, quasi-coherent oscillations occur in a large fraction of CVs, the physical mechanisms producing DNOs, lpDNOs, and QPOs are likely to be of quite general significance to accretion disc physics.  Judging by the time scale and coherence of these oscillations, DNOs and lpDNOs most likely originate in the inner disc or on the white dwarf itself, while the accretion disc must be the source of the QPOs.  The most viable model of DNOs and lpDNOs ascribes them to the interaction of a weak white dwarf magnetic field with the inner accretion flow (a detailed description of this model for DNOs may be found in \citealt{dno2}, hereafter Paper II).  The rapid oscillations therefore deserve attention not only because they are a common property of disc-fed accretion in CVs, but also because they probably hold diagnostic potential in the study of the accretion process and the interaction between discs and magnetic fields.  

The discovery of a correlation between the ratio of time scales of the oscillations in CVs and those in low-mass X-ray binaries (LMXBs) has further renewed interest in DNOs and QPOs (Woudt \& Warner 2002, hereafter Paper I; Paper II; \citealt{p17}; Paper III).  Accretion processes are more readily studied in CVs than in LMXBs---observations of the oscillations in CVs can be obtained with modest-sized ground based telescopes, and may provide insight also into the QPOs found in LMXBs. 

This is the fifth in a series of papers on rapid, quasi-coherent oscillations in CVs; we present more observations of these oscillations in the next four sections, and discuss the results in Section \ref{sec:discuss}.  A more complete account of this work may be found in \citet{thesis}.

\section{Observations}

We obtained high speed photometry of CVs at the Sutherland site of the South African Astronomical Observatory using the 30-, 40-, and 74-in. telescopes and the University of Cape Town charge-coupled device photometer (the UCT CCD), a frame-transfer CCD (\citealt{uctccd} describes the instrument).  Table \ref{tab:obs} gives a log the observations discussed here; for the most part only observations that lead to the detection of oscillations are listed.  In almost all cases the observations were made in white light; with the UCT CCD this gives photometry with an effective wavelength similar to Johnson $V$, but with a very broad bandpass.  This, together with the non-standard flux distribution of CVs, means that magnitude calibration (achieved by observing hot subdwarf and white dwarf standard stars) and extinction corrections are only approximate (accurate to $\approx0.1$~mag).  

We have focused our observational effort on systems with high $\dot{M}$, i.e. nova-like variables and dwarf novae in outburst, but have also observed a few quiescent dwarf novae.

\begin{table*}
 \centering
 \begin{minipage}{150mm}
  \caption{Log of the observations.  The average magnitude for each run is listed in the final column.}
  \label{tab:obs}
  \begin{tabular}{@{}llllllll@{}}
  \hline
Object  &  Run no. &  Date at the start & First integration  & Length (h) & $t_{int}$ (s) &  Telescope  &  $V$ (mag)  \\
        &          &  of the night      & HJD $=2450000.0+$  &            &    &   & \\ 
 \hline
WW Cet    & S7082 &  3  Sept. 2003 & 2886.55197 & 2.78 & 5          & 40-in. & 11.8     \\
BP CrA    & S6912 &  4  Apr.  2003 & 2734.55464 & 2.73 & 6          & 30-in. & 14.3:    \\
          & S7025 & 25  Jul.  2003 & 2846.35335 & 2.55 & 5, 6       & 40-in. & 13.8     \\
          & S7029 & 26  Jul.  2003 & 2847.27017 & 2.78 & 6          & 40-in. & 13.7     \\
          & S7035 & 28  Jul.  2003 & 2849.31426 & 1.92 & 5          & 40-in. & 13.8     \\
          & S7161 & 19  Oct.  2003 & 2932.23729 & 2.23 & 5, 6       & 40-in. & 14.0     \\
BR Lup    & S6911 &  4  Apr.  2003 & 2734.41199 & 3.37 & 6          & 30-in. & 14.5     \\
          & S6914 &  7  Apr.  2003 & 2737.42124 & 2.62 & 6          & 30-in. & 14.3     \\
          & S7015 & 22  Jul.  2003 & 2843.20290 & 5.28 & 7, 8       & 40-in. & 15.7     \\
HP Nor    & S6837 &  5  Mar.  2003 & 2704.46186 & 4.23 & 6          & 30-in. & 14.0     \\
          & S7028 & 26  Jul.  2003 & 2847.20273 & 1.48 & 6          & 40-in. & 13.1     \\
AG Hya    & S6907 &  3  Apr.  2003 & 2733.29331 & 4.12 & 6          & 30-in. & 14.7:    \\
V1193 Ori & S6648 & 25  Nov.  2002 & 2604.41708 & 1.38 & 4, 8       & 40-in. & 14.0     \\
          & S6652 & 26  Nov.  2002 & 2605.50435 & 2.70 & 6          & 40-in. & 14.0     \\
          & S6655 & 28  Nov.  2002 & 2607.40756 & 5.15 & 6          & 40-in. & 14.0     \\
          & S6659 & 29  Nov.  2002 & 2608.46563 & 3.47 & 5, 6       & 40-in. & 13.8     \\
          & S6663 & 30  Nov.  2002 & 2609.47053 & 2.02 & 6          & 40-in. & 14.0     \\
          & S6668 &  1  Dec.  2002 & 2610.40801 & 3.12 & 5          & 40-in. & 13.9     \\
          & S6675 & 17  Dec.  2002 & 2626.54630 & 1.32 & 8          & 40-in. & 14.8     \\
          & S6689 & 23  Dec.  2002 & 2632.32129 & 4.30 & 5          & 40-in. & 13.9     \\
          & S6692 & 24  Dec.  2002 & 2633.48646 & 2.95 & 4, 5       & 74-in. & 14.0     \\
          & S6694 & 25  Dec.  2002 & 2634.32814 & 4.95 & 4, 5       & 74-in. & 14.0     \\
          & S6696 & 26  Dec.  2002 & 2635.48094 & 2.85 & 5          & 74-in. & 14.0     \\
          & S6698 & 27  Dec.  2002 & 2636.33502 & 3.55 & 5          & 74-in. & 14.0     \\
RU Peg    & S7009 & 16  Jul.  2003 & 2837.53233 & 4.52 & 5, 4, 2    & 40-in. & 9.0$^a$  \\
          & S7010 & 18  Jul.  2003 & 2839.56055 & 3.17 & 5, 4       & 40-in. & 9.2$^a$  \\
          & S7011 & 19  Jul.  2003 & 2840.53717 & 3.93 & 5, 4, 3, 1 & 40-in. & 9.4$^a$  \\ 
          & S7012 & 20  Jul.  2003 & 2841.53055 & 3.87 & 5, 4, 2    & 40-in. & 9.5$^a$  \\ 
          & S7013 & 21  Jul.  2003 & 2842.53382 & 1.88 & 4          & 40-in. & 9.7$^a$  \\ 
          & S7016 & 22  Jul.  2003 & 2843.43586 & 4.13 & 5, 4       & 40-in. & 9.8$^a$  \\
          & S7021 & 23  Jul.  2003 & 2844.53042 & 4.02 & 6, 5       & 40-in. & 10.1$^a$ \\  
          & S7026 & 25  Jul.  2003 & 2846.46600 & 1.72 & 4          & 40-in. & 10.7$^a$ \\
HX Peg    & S7162 & 19  Oct.  2003 & 2932.34182 & 3.68 & 5          & 40-in. & 14.8     \\
V426 Oph  & S6935 & 18  May   2003 & 2778.49935 & 1.57 & 6          & 40-in. & 12.1     \\
V1159 Ori & S6649 & 25  Nov.  2002 & 2604.48413 & 3.07 & 4, 5       & 40-in. & 12.9     \\
          & S6784 & 21  Feb.  2003 & 2692.25652 & 0.88 & 5          & 40-in. & 12.9     \\
X Leo     & S6742 &  5  Feb.  2003 & 2676.37284 & 0.78 & 5          & 40-in. & 14.0     \\
          & S6921 & 14  May   2003 & 2774.19907 & 1.28 & 5          & 40-in. & 12.9     \\
          & S6923 & 14  May   2003 & 2774.32065 & 1.50 & 5          & 40-in. & 12.9     \\
          & S6927 & 15  May   2003 & 2775.19494 & 4.17 & 4, 5, 6    & 40-in. & 13.2     \\
          & S6931 & 17  May   2003 & 2777.19303 & 1.92 & 5          & 40-in. & 14.5     \\
          & S6932 & 18  May   2003 & 2778.19122 & 2.80 & 7          & 40-in. & 14.9     \\
          & S7293 & 16  Mar.  2004 & 3081.24386 & 1.45 & 5          & 40-in. & 12.7     \\
          & S7295 & 17  Mar.  2004 & 3082.24245 & 5.58 & 5, 6       & 40-in. & 13.2     \\
          & S7297 & 19  Mar.  2004 & 3083.24244 & 3.68 & 6          & 40-in. & 14.9     \\
V893 Sco  & S6942 & 19  May   2003 & 2779.47549 & 1.02 & 6          & 40-in. & 13.0$^b$:\\
          & S6945 & 20  May   2003 & 2780.41393 & 5.98 & 7          & 30-in. & 13.9$^b$ \\
          & S6949 & 21  May   2003 & 2781.39698 & 3.95 & 7          & 30-in. & 15.1$^b$ \\
          & S6953 & 23  May   2003 & 2783.38547 & 2.97 & 6          & 30-in. & 14.0$^b$ \\
WX Hyi    & S6463 & 10  Jul.  2002 & 2466.48583 & 3.30 & 6          & 30-in. & 15.0     \\ 
          & S7192 & 24  Dec.  2003 & 2998.32532 & 1.62 & 5          & 40-in. & 14.6     \\   
\hline
\end{tabular}
Notes: $t_{int}$ is the integration time; `:' denotes an uncertain value; $^a$ $U$ magnitude; $^b$ average magnitude outside of eclipse. \hfill
\end{minipage}
\end{table*}

\section{Systems with no previous detections of oscillations}

\subsection{WW Cet}
\begin{figure*}
 \includegraphics[width=140mm]{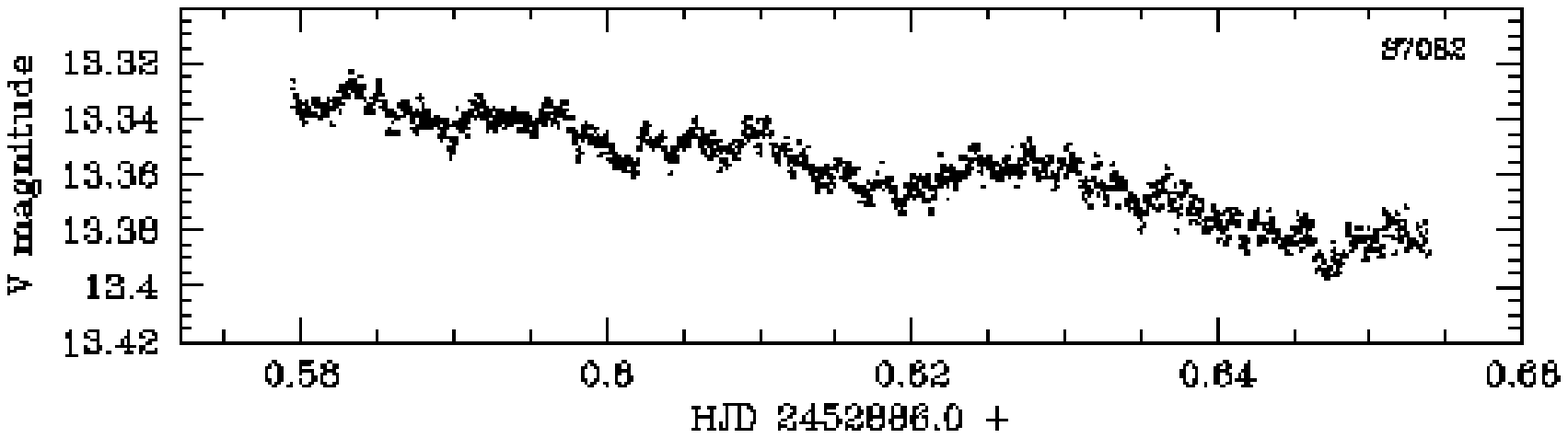}
 \caption{Part of the light curve of WW Cet on 3 September 2003, when the system  was on the decline from outburst.  The data are corrected to first order for atmospheric extinction.}
 \label{fig:s7082lc}
\end{figure*}

There is some uncertainty as to what type of dwarf nova WW Cet is (e.g.\ \citealt{wwcetty}; \citealt{wwcetper}); it has an orbital period ($P_{orb}$) of 4.219~h, a mean quiescent $V$ magnitude of 13.9 (but can get as faint as 15.7~mag) and reaches 9.3 mag in outburst \citep{wwcetper}.  WW Cet was observed during two outbursts and, on one occasion, in quiescence; rapid oscillations were detected on only one night.  Fig. \ref{fig:s7082lc} is the light curve for run S7082, which shows WW Cet fading from outburst at an average rate of 0.029~mag/h. 

\begin{figure*}
 \includegraphics[width=140mm]{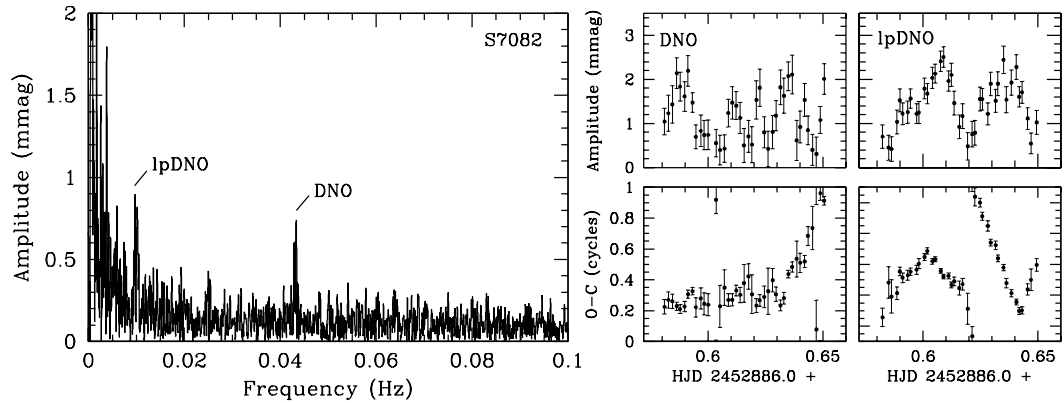}
 \caption {Left: The Fourier transform of the light curve shown in Fig. \ref{fig:s7082lc}.  An lpDNO at $\sim$0.0097~Hz (103~s) is present at the same time as a DNO at $\sim$0.043~Hz (23.1~s).  The light curve was prewhitened to remove a telescope drive error at 40 sidereal seconds.  Right: Amplitude and $O-C$ variation of the DNO and lpDNO in run S7082.  The DNO phase is calculated relative to a period of 23.08~s, that of the lpDNO relative to a period of 102.7~s.  The DNO period increases systematically, while the lpDNO period changes quite abruptly at around HJD 2452886.603, and again at around HJD 2452886.643.}
 \label{fig:s7082ftomc}
\end{figure*}

Peaks at $\sim$0.0097~Hz (corresponding to 103~s) and $\sim$0.043~Hz (23.1~s) in the Fourier transform of this light curve (left hand panel of Fig. \ref{fig:s7082ftomc}) indicate the presence of an lpDNO and DNO.  Both the DNO and lpDNO peaks contain some noisy structure.  Amplitude and $O-C$ phase diagrams show this to be caused by changing periods and amplitudes---see the right hand side of Fig. \ref{fig:s7082ftomc}.  $P_{DNO}$ shows the usual increase with decreasing accretion luminosity, while the DNO amplitude varies erratically.  The lpDNO period does not change systematically during the observation, rather, it changes abruptly from $\sim$104~s to $\sim$98.6~s at HJD 2452886.603, and increases again at HJD 2452886.643.  These changes in period are not accompanied by any remarkable changes in system luminosity (see Fig. \ref{fig:s7082lc}) or oscillation amplitude (see the right most upper panel of Fig. \ref{fig:s7082ftomc}).  There does not appear to be any obvious correlation between the phase and amplitude behaviour of the DNO and lpDNO.  The classification of the $\sim$100~s oscillation as an lpDNO, rather than a QPO is based on its period being around 4 times that of the DNO period ($P_{lpDNO}/P_{DNO}=4.27$).

There is a sharp spike in the Fourier transform at 3.802~mHz (263~s), at larger amplitude than the surrounding noise.  The spike is probably caused by a QPO, but the modulation responsible for it is not obvious in the light curve.

\subsection{BP CrA}
BP CrA is a Z Cam type dwarf nova with visual magnitude varying between 15.9 and 13.5, and an outburst recurrence time of 13.5~d (\citealt{cra1},b); there is no orbital period known for this object.

\begin{figure*}
 \includegraphics[width=140mm]{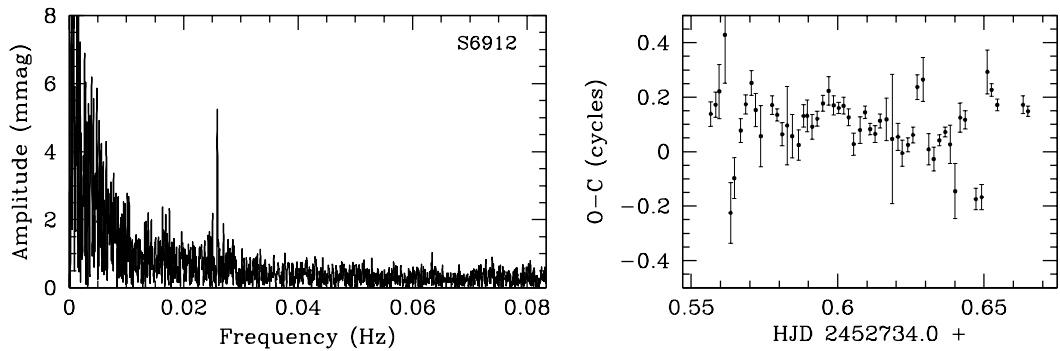}
 \caption {The Fourier transform of a light curve of BP Cra (run S6912; left hand panel), and $O-C$ diagram of the DNO (right hand panel).  The DNO period is 38.64~s (0.02588~Hz).  The light curve was prewhitened to correct for the 30-in. telescope's drive error of 120 sidereal seconds.  The $O-C$ values are calculated relative to a fixed period of 38.64~s, using about 8 cycles per point, and 50\% overlap between consecutive points, so that every second point is independent.}
 \label{fig:s6912ft}
\end{figure*}

Observations of BP CrA made during outburst on 4 April 2003 (run S6912) contained a strong DNO.  The system was subsequently observed on nine more nights, but the DNO was never detected again.  Fig. \ref{fig:s6912ft} displays the Fourier transform of run S6912, which shows that the DNO period is 38.64~s; the amplitude (refined by applying a non-linear least squares fit) is 5.8~mmag.  Also shown in Fig. \ref{fig:s6912ft} is the $O-C$ phase diagram of the DNO, which shows variability typical of a DNO, with the period initially longer than the test period of 38.64~s (indicated by $O-C$ sharply increasing with time), and then, at around HJD 2452734.575, changing to be more consistent with the test period. 

\begin{figure*}
 \includegraphics[width=140mm]{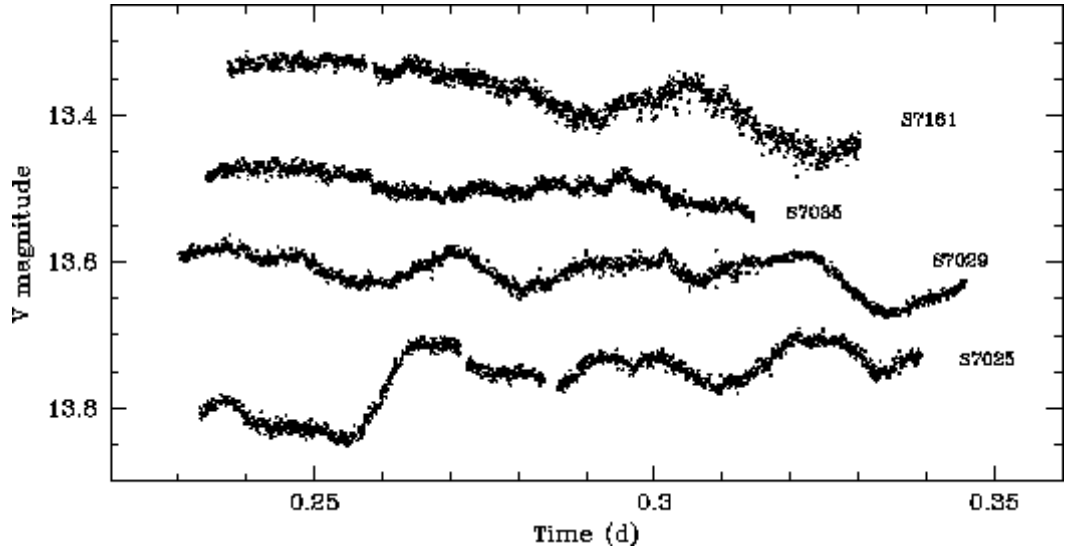}
 \caption {Four light curves of BP CrA, arbitrarily displaced along the horizontal axis.  The S7029, S7035, and S7161 light curves are shifted vertically by $-0.07$, $-0.35$, and $-0.63$~mag for display purposes.}
 \label{fig:bpcralcs}
\end{figure*}

Since there is no high speed photometry of BP CrA published, we display a number of light curves in Fig. \ref{fig:bpcralcs}.  The data set did not reveal the orbital period.

\subsection{BR Lup}
BR Lup is an SU UMa type dwarf nova \citep{luptype}, with an orbital period of (probably) 1.909~h (\citealt{luptype}; \citealt{lupper}).  

\begin{figure*}
 \includegraphics[width=140mm]{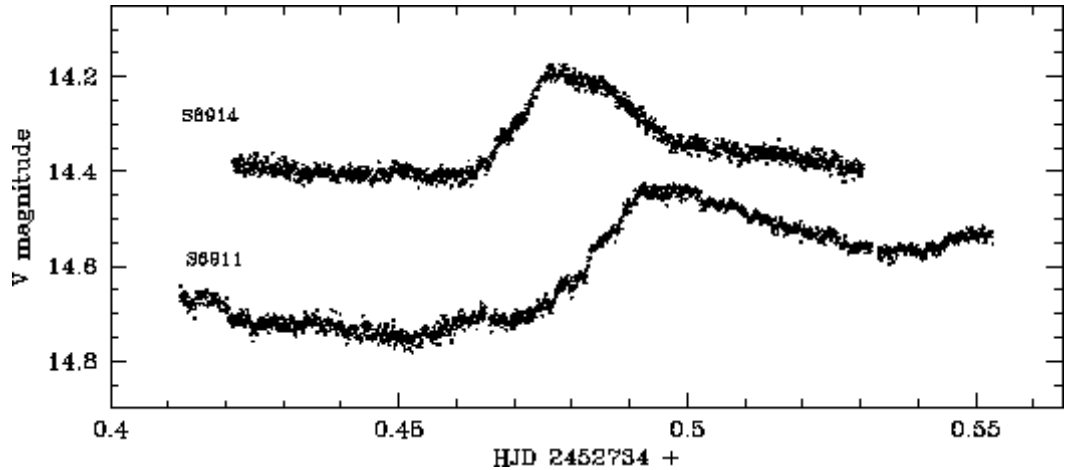}
 \caption{Light curves of BR Lup on 4 and 7 April 2003.  On 4 April (run S6911) the system was clearly still rising into superoutburst.  The 7 April data (S6914) are translated along the horizontal axis by $-3.0$~d; no vertical shift was applied.  The large amplitude modulations are superhumps.}
 \label{fig:brluplc}
\end{figure*}

\begin{figure*}
 \includegraphics[width=140mm]{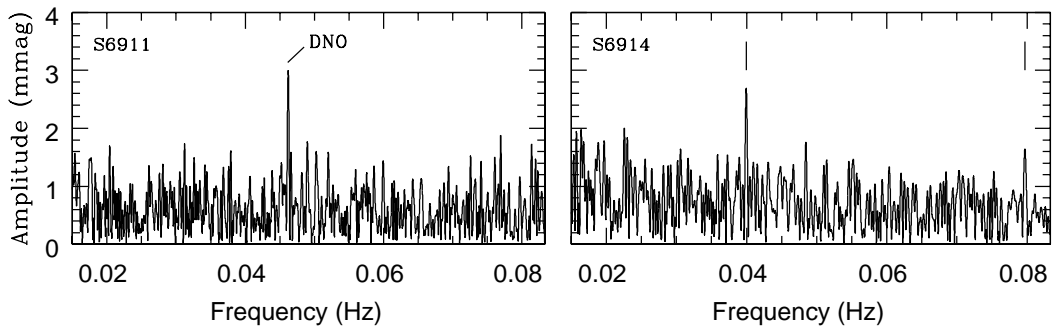}
 \caption{The high frequency end of Fourier transforms of the two BR Lup light curves.  Left: The first 1.3~h of run S6911, with a DNO at 21.65~s marked.  Right: The central third of run S6914.  Vertical bars mark peaks at 25.038~s and 12.548~s (12.548~s is within 0.231\% of where the first harmonic of 25.038~s is).}
 \label{fig:brlupfts}
\end{figure*}

\begin{figure*}
 \includegraphics[width=140mm]{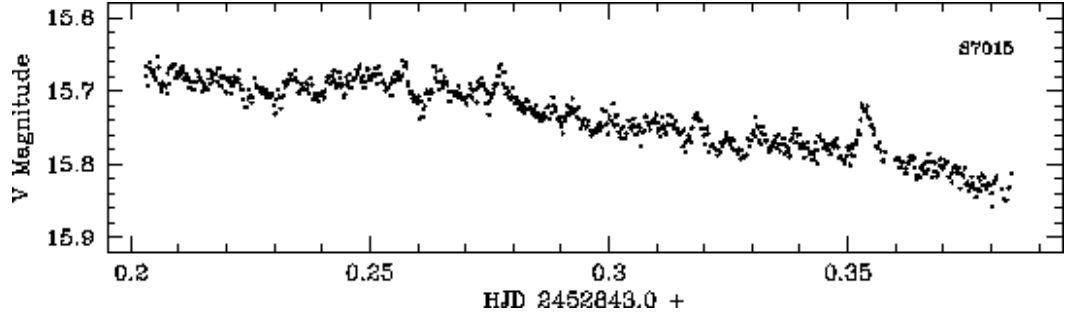}
 \caption{The light curve of BR Lup on 22 July 2003 (run S7015) binned to a third of its original time resolution.  These data show a QPO with a period of $\sim$650~s.  The light curve contains no higher frequency signals.}
\label{fig:s7015lc}
\end{figure*}

Fig. \ref{fig:brluplc} shows the light curves of run S6911 (taken on the rise to supermaximum) and S6914.  A DNO was detected in run S6911; the left hand panel of Fig.~\ref{fig:brlupfts} is the Fourier transform of the first  1.3~h of this light curve, with the DNO at 21.65~s.  The Fourier transform of the central third of run S6914 (right hand panel of Fig. \ref{fig:brlupfts}) shows the presence of a modulation at 25.04~s; since the system was brighter on this night than during run S6911, one would expect $P_{DNO}$ to have decreased.  However, the 25.04~s signal in S6914 is accompanied by a spike (only just visible above the noise) at the first harmonic (12.55~s).  A non-sinusoidal profile is characteristic of a DNO reprocessed off a QPO structure (\citealt{p19}; Paper II), but no QPO was seen in either of these light curves.  Note that, depending on the geometry, it is possible to see the reprocessed DNO, without the QPO itself being visible.  Run S7015, taken during a different outburst, contains a QPO with a period near 650~s, Fig. \ref{fig:s7015lc} shows the light curve at 21~s time resolution; the QPO is clearly visible.  


\subsection{HP Nor}
\begin{figure*}
 \includegraphics[width=140mm]{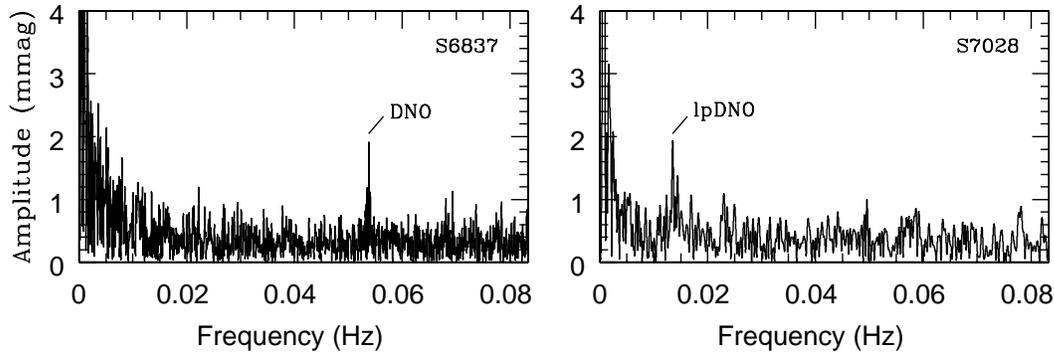}
 \caption {Left: The Fourier transform of the first 2.1~h of the light curve of HP Nor on 5 March 2003 (run S6837).  The period determined from fitting a sine function to the DNO by non-linear least squares is 18.58~s.  Right: The Fourier transform of a 0.76~h section of the light curve of HP Nor on 26 July 2003 (run S7028) showing an oscillation at 74.01~s, which we interpret as an lpDNO.  The lpDNO was present only in this short section of data.}
 \label{fig:hpfts}
\end{figure*}

\begin{figure*}
 \includegraphics[width=140mm]{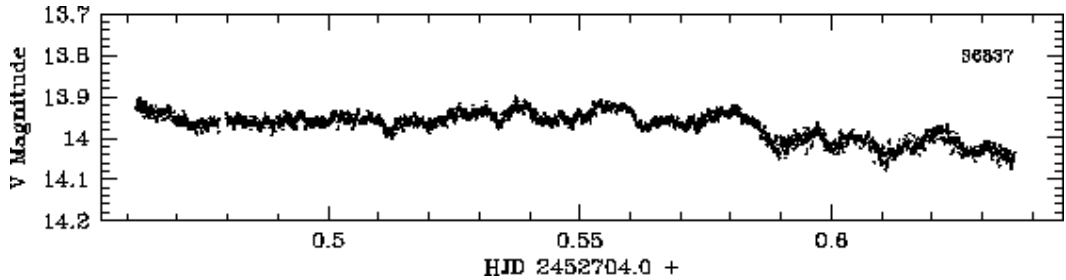}
 \caption {The light curve of HP Nor on 5 March 2003, corrected to first order for atmospheric extinction.  The quality of photometry  deteriorates towards the end of the run, despite decreasing airmass, because of worsening seeing.}
 \label{fig:s6837lc}
\end{figure*}

HP Nor is a Z Cam star \citep{hptype} with a maximum brightness of $V=12.6$, and minimum of $V=16.4$ \citep{hprange}.  Both a DNO and an lpDNO were detected in HP Nor, but in separate outbursts\footnote{The 35.2~s DNO period listed in W04 is incorrect.}.  Fig. \ref{fig:hpfts} displays Fourier transforms of parts of the light curves of run S6837 and S7028, showing a DNO and an lpDNO respectively.  The DNO has a best-fit period of 18.58~s, and amplitude of 1.9~mmag; it persisted at detectable amplitude for the entire run.  The lpDNO appeared only briefly; in the $\sim$0.76~h section of data where it was most prominent, it has a period of 74.01~s, and an amplitude of 2.0~mmag.  We display our longest light curve of HP Nor (run S6837) as Fig. \ref{fig:s6837lc}, since there is no high speed photometry of this system in the literature.  

\subsection{AG Hya}
Very little is known about AG Hya; it is classified as a U Gem star \citep{hptype}, and has been observed to vary in the range $V=19.2$ to $V=13.8$ (\citealt{szk87}; \citealt{aghya}).   No high speed photometry of AG Hya has been reported, and the orbital period is not known.

\begin{figure*}
 \includegraphics[width=140mm]{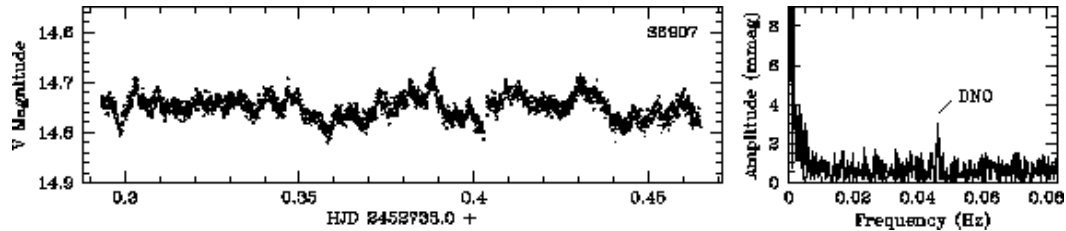}
 \caption {Left: The light curve of AG Hya on 3 April 2003, after being corrected for atmospheric extinction.  Around 20 cycles of a QPO with a period of 683~s are directly visible in this light curve.  Right: The Fourier transform of a $\sim$0.9~h section of the light curve on the left, the DNO is at 46.40~mHz (21.55~s).}
 \label{fig:aghyalcft}
\end{figure*}

We collected a total of 13~h of observations of this system; one of our runs contains a QPO with a period of 683~s, as well as a 21.55~s DNO.  Fig. \ref{fig:aghyalcft} shows the light curve, in which the QPO is quite clear, together with the Fourier transform of a $\sim$0.9~h section of the data.  The DNO amplitude is sufficiently large for it to appear above the noise only in this short section of data, and it was never observed again.  The low amplitude and brief presence make a more detailed investigation of the DNO impossible.

\subsection{V1193 Ori}
\begin{figure*}
 \includegraphics[width=140mm]{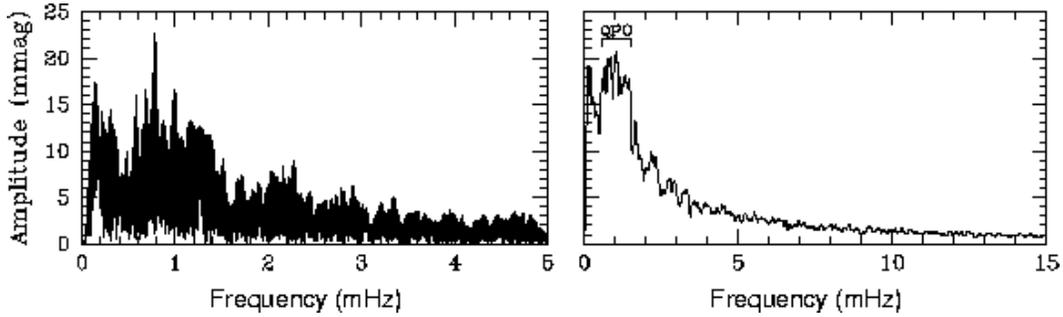}
 \caption {Left: The Fourier transform of the first six runs on V1193 Ori (taken over seven nights; see Table \ref{tab:obs}) combined.  The data were prepared by subtracting a linear trend from each light curve individually before calculating the Fourier transform.  The maximum amplitude is at 0.7834~mHz (1276~s).  Right:  The average Fourier transform of the whole V1193 Ori data set.  The QPO appears as a hump of power at frequencies between 0.580~mHz (1720~s) and 1.54~mHz (649~s), superimposed on the smoothly rising red noise.}
 \label{fig:v1193fts}
\end{figure*}

V1193 Ori was discovered as a rapidly varying blue star by \citet{v1193}, and classified as a UX UMa type nova-like variable by \citet{v1193type}; the spectroscopically determined orbital period is 3.96~h \citep{v1193per}.  What appears to be rapid, large amplitude flickering is evident in every published light curve of this object (\citealt{v1193type}; \citealt{v1193phot}; \citealt{papad}).  Our light curves of V1193 Ori are dominated by kilosecond QPOs, rather than flickering, although they look very similar to those of \citet{v1193type} and \citet{v1193phot}.  Despite a period that changes from night to night, the QPO stands out in the Fourier transform of the first six runs combined (see the left hand panel of Fig. \ref{fig:v1193fts}), which implies unusually large coherence.  The average Fourier transform of all the observations of V1193 Ori shows power in excess of red noise at frequencies between 0.580~mHz (1720~s) and 1.54~mHz (649~s); this is displayed in the right hand panel of Fig. \ref{fig:v1193fts}.

\section{Previously studied systems}

The CVs discussed in this section are already known to possess rapid oscillations; we highlight only those of our observations that reveal new aspects of the phenomenology, or additional classes of oscillations.

\subsection{RU Peg}
RU Peg is a U Gem type dwarf nova with outbursts of about 3 mag recurring every 75--85~d and has an orbital period of 8.990~h \citep{rupegper}.  This is the CV in which QPOs were first recognised.  \citet{p11} found a QPO with a period near 50~s present at the same time as a signal at $\sim$12~s.  The 12~s signal was detected on four nights and was believed to be a DNO; it was suggested in Paper III that the 12~s signal seen in RU Peg might have been a beat between a DNO at 2.97~s and the 4~s integration time used by Patterson et al.  

\begin{figure*}
 \includegraphics[width=140mm]{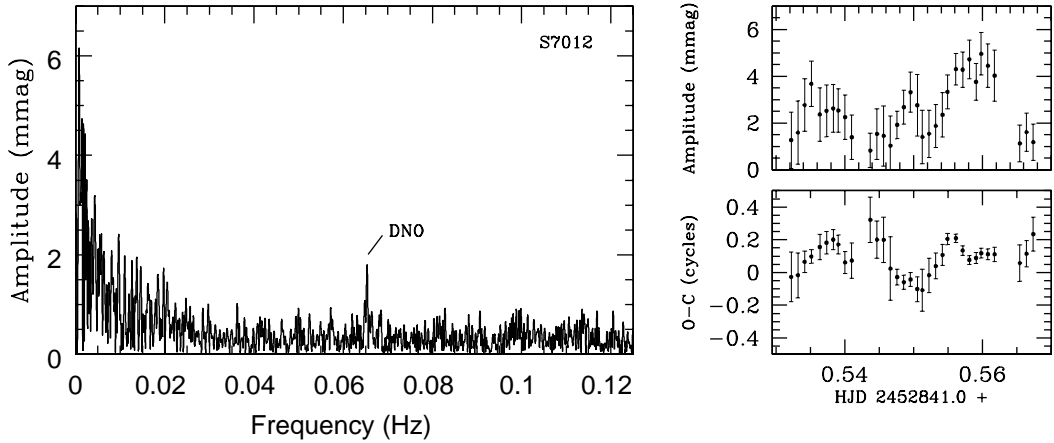}
 \caption {Left: The Fourier transform of a 58~min.\ section of obseravtions of RU Peg (run S7012; taken at 4~s time resolution) after a second order polynomial trend was removed.  The DNO period is 15.30~s.  There is no evidence of a QPO in these data. Right: The amplitude and $O-C$ diagram of the DNO in run S7012 relative to a fixed period of 15.03~s using $\sim$16 DNO cycles per point, and with 67\% overlap between consecutive points.  A few points of very low amplitude are not shown.  Notice that the amplitude is large for only about 12~min., and that it becomes very low towards the end of this part of the observation.}
 \label{fig:s7012ftomc}
\end{figure*}

In an attempt to find out if RU Peg indeed has DNOs at very short periods, we observed the July 2003 outburst.  Although we detected oscillations at 13.94~s and 15.03~s in run S7010 and S7012 respectively, we failed to discern whether these were caused by DNOs at shorter periods.  In both instances we changed the integration time as soon as we found a clear signal, but the oscillations did not persist for long enough to determine whether the periods changed when the integration time was adjusted.  Fig.~\ref{fig:s7012ftomc} illustrates the oscillation in S7012.  

The lowest frequency signals that could beat with the integration time to produce the peak seen in run S7010 are at 3.68~s and 7.80~s.  For run S7012 these aliases have periods of 3.16~s and 5.45~s.  It seems very unlikely that higher frequency DNOs would produce signals below the Nyquist frequency at large enough amplitude to be detected.  Note e.g.\ that if a DNO was present in S7012 at 2.31~s (the alias corresponding to $2/t_{int}-1/15.03$~s, where $t_{int}$ is the integration time) the observed amplitude would be a factor of (sin\,$x$)/$x$ smaller than the true amplitude, where $x=\pi t_{int}/P_{DNO}$.  The observed amplitude was, for a while, at least as large as 3~mmag (see Fig. \ref{fig:s7012ftomc}).  This would imply, if the observed signal was a beat, that the true amplitude was greater than 22~mmag.  DNOs at such large amplitude are rare, but not unheard of (see e.g.\ \citealt{typsa}; Paper I).  Still, the amplitudes of the $\sim$14 and $\sim$15~s oscillations are typical of DNOs or lpDNOs whereas that of the 12~s signal seen by Patterson et al. was unusually low.  A further constraint is that, since the observations were made on the decline from outburst, the DNO period should be longer in S7012 than in S7010.  This leaves the possibility of a DNO at 3.68~s in S7010 and at 5.45~s in S7012.  

Arguing that $P_{QPO}/P_{lpDNO}\approx 4$ one could conclude, for $P_{QPO}\approx 50$~s, that the 12, 14 and 15~s signals are all lpDNOs (Paper III). This argument fails for the QPO periods observed later in the outburst.  Assuming that the 14 and 15~s periods are not beats with DNOs above the Nyquist frequency, the period changed by more than 7\% while the system faded by 0.32~mag in $U$.  This period change is quite normal for a DNO, but the period of an lpDNOs is usually less sensitively dependent on luminosity  (of course, lpDNOs are not yet a very well studied phenomenon; furthermore, in Section \ref{sec:xleo}, we present an example of a system in which $P_{lpDNO}$ does depend sensitively on luminosity).  It is also possible that, e.g., the observed period in S7010 was a beat between a higher frequency signal and the integration time, while that in S7012 was the true oscillation period.  We tentatively classify the 14 and 15~s signals as lpDNOs.

\begin{figure*}
 \includegraphics[width=140mm]{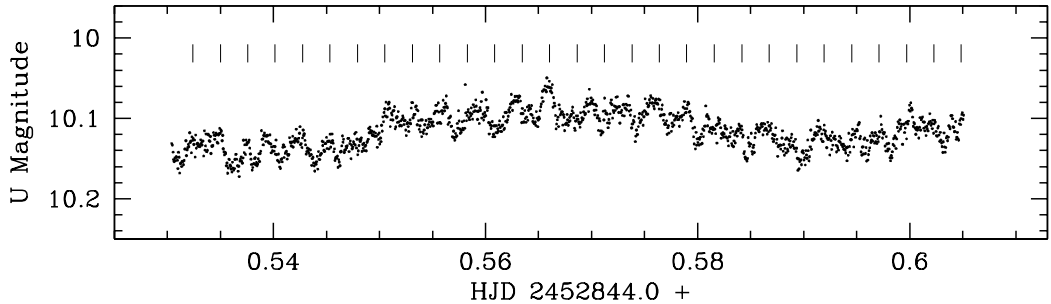}
 \caption {The first 107.5 min.\ of the light curve of RU Peg on 23 July (run S7021). Vertical bars along the top of the graph are equally spaced at 223.6 s intervals, which is the mean QPO period in this part of the data.  Non-linear least squares gives a best-fit amplitude of 5.649~mmag for this QPO.}
 \label{fig:s7021lc}
\end{figure*}

\begin{figure*}
 \includegraphics[width=140mm]{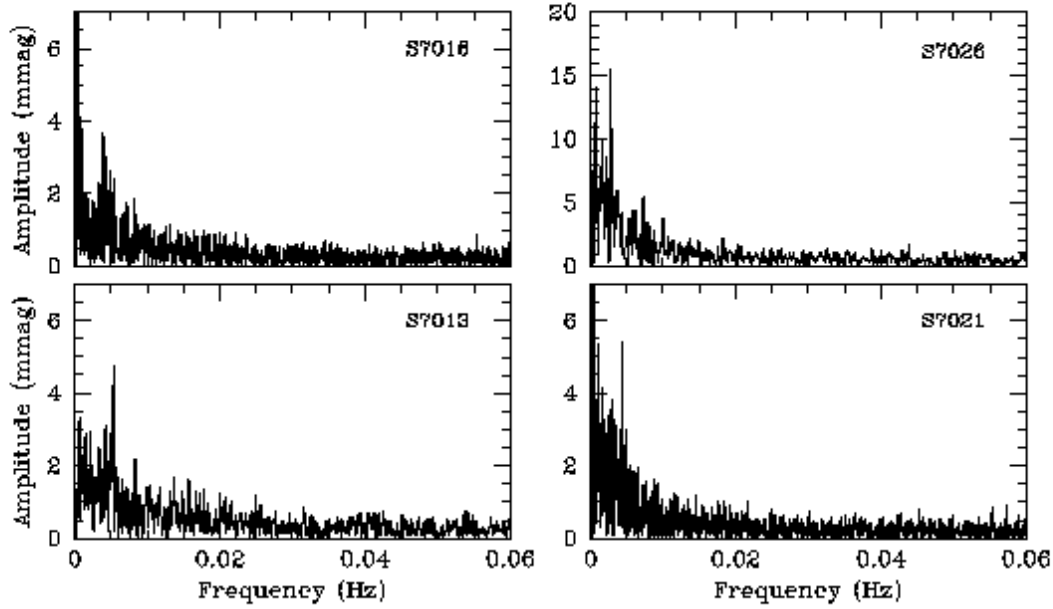}
 \caption {Fourier transforms showing QPOs in RU Peg on four nights.  The QPOs are much more coherent here than earlier in the outburst and appear as sharp peaks (or, in the case of S7016, a narrow band of power).}
 \label{fig:rupegfts}
\end{figure*}

QPOs were detected in most of the observations we obtained of RU Peg, and showed a large period evolution over the decay to quiescence.  Table~\ref{tab:rupeg} lists the average QPO period, as well as the best-fit amplitude, for every night on which QPOs were present.  An example of a light curve of RU Peg with an obvious QPO is given in Fig.~\ref{fig:s7021lc}; some of the Fourier transforms are shown in Fig.~\ref{fig:rupegfts}.  By 26 July 2003 RU Peg was still above quiescent brightness but only flickering.

\begin{table}
 \caption{The average nightly $U$ magnitude, together with QPO period and amplitude measured in RU Peg.}
 \label{tab:rupeg}
 \begin{tabular}{@{}llll@{}}
 \hline
Run no. & $U$ (mag) & $P_{QPO}$ (s) & Amplitude (mmag) \\
 \hline
S7009 &  9.0 & $\sim$56 & 2.1  \\
S7010 &  9.2 & $\sim$61 & 2.2  \\
S7011 &  9.4 &      149 & 3.1  \\
S7013 &  9.7 &      186 & 4.8  \\
S7016 &  9.8 &      257 & 3.5  \\
S7021 & 10.1 &      223 & 5.6  \\
S7026 & 10.7 &      360 & 15.4 \\
 \hline
 \end{tabular}
\end{table}

\subsection{HX Peg}
HX Peg is a Z Cam type dwarf nova with $V$ magnitude varying in the range 16.6 to 12.9, outbursts recurring every 14 to 21~d \citep{hxpegtype}, and an orbital period of 4.82~h \citep{hxpegper}.  

\begin{figure*}
 \includegraphics[width=140mm]{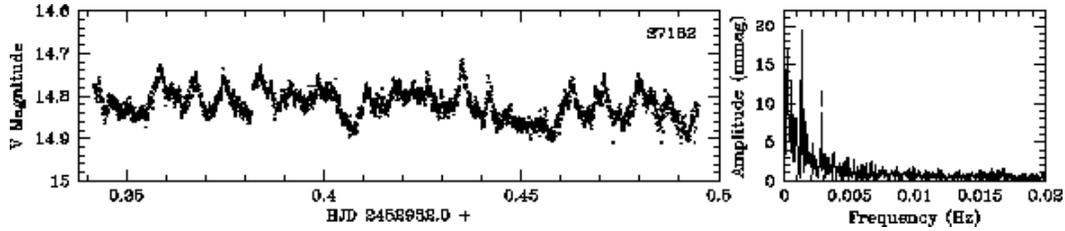}
 \caption {Left: The light curve of run S7162 on HX Peg, dominated by a kilosecond QPO with a mean period of 746~s and amplitude of 19.5~mmag.  This light curve (binned to a quarter of its original time resolution) appeared as fig. 14 of W04.  Right: The low frequency end of the Fourier transform of this light curve, showing, in addition to very large power at 1.34~mHz (746~s), the presence of a normal QPO at a period of 347~s (marked by a vertical bar at 2.88~mHz).  The normal QPO, despite an amplitude of 8.78~mmag, is almost invisible in the light curve, because it is drowned out by the kilosecond QPO.}
 \label{fig:s7162lcft}
\end{figure*}

We observed the system frequently during the 2002 and 2003 seasons, and detected the full set of kilosecond QPOs, QPOs, lpDNOs, and DNOs (see Paper III).  One light curve is of particular interest: run S7162 taken on the decline from outburst is the only example of the coexistence of a normal QPO and a kilosecond QPO (see Fig. \ref{fig:s7162lcft}).  The normal QPO here is at 347~s, while the much more noticeable kilosecond QPO has a period of 746~s.  

\subsection{V426 Oph}

V426 Oph is a Z Cam star with visual magnitude ranging from 10.9 to 13.4, an outburst cycle of about 22~d, and an orbital period of 6.847~h \citep{hess88}.  \citet{szk86} detected a 1~h period in an \textit{EXOSAT} X-ray observation, and proposed that this is the white dwarf spin period, and therefore that V426 Oph is an intermediate polar.  \citet{n+w89} and \citet{hell90} reanalysed the \textit{EXOSAT} data, and found that the 1~h period was, in fact, not statistically significant.  Furthermore, in their optical photometry and spectroscopy, \citet{hell90} find no evidence to suggest that V426 Oph is an intermediate polar.  \citet{szk90} report the presence of a $\sim$28~min.\ QPO in \textit{Ginga} data of this system, but a glance at their Fourier transforms (fig. 7) shows that any power present near 0.0006~Hz is no different from the red noise expected from random flickering.  The latest addition to the confusion is the detection by \citet{v426} of a 4.2~h modulation and its first harmonic in \textit{Chandra} observations of V426 Oph; they also reanalysed \textit{Ginga} \citep{szk90} and \textit{ROSAT} \citep{r94} data taken in 1988, and show that these data are just compatible with the 4.2~h or 2.1~h periods, but certainly do not demonstrate that ``...the 4.2~h period or its harmonic are stable and persistent in X-ray/optical data from 1988 to 2003.''.  \citet{v426} propose that the 4.2~h period may be identified with the rotation of the white dwarf, but point out that this interpretation is physically unreasonable.

\begin{figure*}
 \includegraphics[width=140mm]{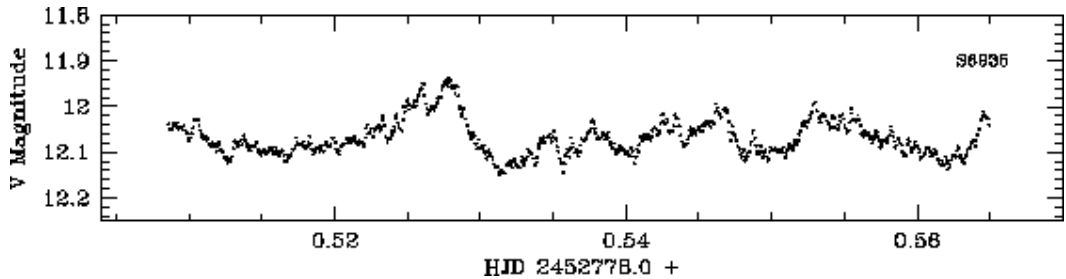}
 \caption {The light curve of V426 Oph on 18 May 2003 (run S6935), corrected to first order for atmospheric extinction.  The rapid flickering disguises a quasi-coherent 73 s modulation---see Fig. \ref{fig:s6935ftomc}.}
 \label{fig:s6935lc}
\end{figure*}

\begin{figure*}
 \includegraphics[width=140mm]{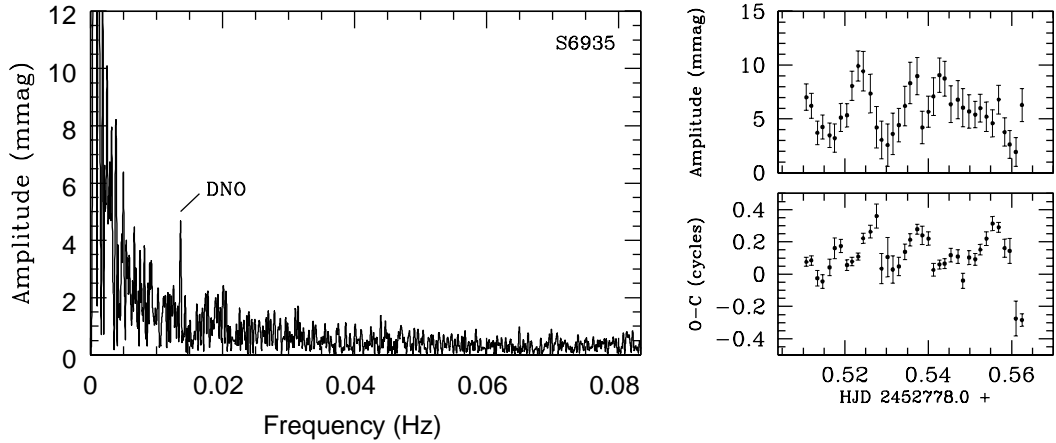}
 \caption {Left: The Fourier transform of the light curve shown in Fig. \ref{fig:s6935lc}, the narrow spike at 0.01365~Hz (73.26~s) is interpreted as a DNO or lpDNO.  Right: Amplitude and $O-C$ phase diagrams of the 73~s modulation.  Each point represents a non-linear least squares fit of a sine function with period 73.26~s to $\sim$5 cycles of the modulation; there is 67\% overlap between consecutive points.}
 \label{fig:s6935ftomc}
\end{figure*}

We obtained high speed optical photometry of V426 Oph on eight nights in 2003.  No single observation was long compared to any of the profusion of periods reported for this system, and no sign of any persistent period was found in the data set.  The only modulation to surface above the flickering was a signal at $\sim$73~s in run S6935, taken on 18 May 2003 (see Table~\ref{tab:obs}).  Fig. \ref{fig:s6935lc} shows the relevant light curve.  A few of the most prominent cycles of the 73 s oscillation are directly visible, but the the Fourier transform (left hand panel of Fig. \ref{fig:s6935ftomc}) makes the signal more evident.  Also displayed in Fig. \ref{fig:s6935ftomc} is the amplitude and phase variability of this modulation.  The $O-C$ diagram shows that the  73~s oscillation has fairly high coherence, but is not strictly periodic (note that in this diagram only every third point is independent); from the amplitude diagram can be seen that the oscillation reached amplitudes as large as 10~mmag.

For the moment we classify the 73~s oscillation as a DNO; the detection of a shorter period in V426 Oph may in the future change this.  Until a similar time scale is observed in V426 Oph, one cannot have absolute confidence in the reality of the DNO.  However, should the oscillation (and its quasi-periodic nature) be confirmed, the favoured model for DNOs (and lpDNOs) excludes V426 Oph from being an intermediate polar.

\subsection{V1159 Ori}
There is a group of SU UMa stars with unusually short recurrence times of normal and superoutbursts, sometimes referred to as ER UMa stars.  V1159 Ori is such a system (\citealt{nkmh95}; \citealt{rht95}) it has a $\sim$4~d recurrence period for normal outbursts \citep{jc92}, a $\sim$45~d supercycle \citep{rht95}, and $P_{orb}=1.49227$~h (\citealt{jc92}; \citealt{ttbr97}).  The most comprehensive photometric study of this system is that by \citet{p92}; they detected DNOs with a  period evolving from 29~s to 34~s over 7 nights of the decline from superoutburst.  

\begin{figure*}
 \includegraphics[width=140mm]{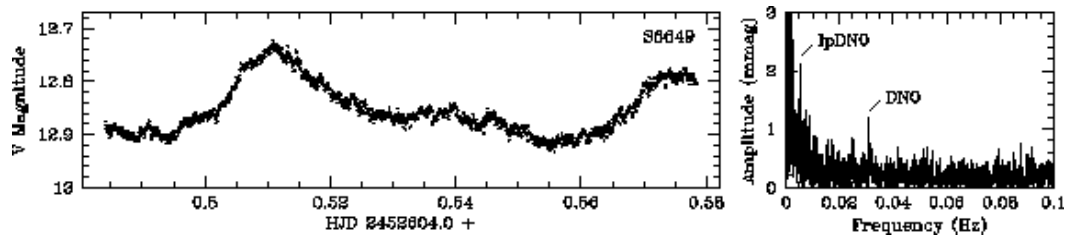}
 \caption {Left: The light curve of a 2.3~h section of run S6649 on V1159 Ori, corrected to first order for atmospheric extinction.  The large amplitude modulation is a superhump.  Right: The Fourier transform of the light curve shown on the left. The DNO at 0.03097~Hz (32.28~s) and lpDNO at 0.00565~Hz (177~s) are indicated.}
 \label{fig:s6649lcft}
\end{figure*}

We detected rapid oscillations in V1159 Ori on 25 November (run S6649), during a superoutburst.  The left hand panel of Fig. \ref{fig:s6649lcft} displays part of this light curve.  The Fourier transform (right hand panel of Fig. \ref{fig:s6649lcft}) shows a DNO at 32.28~s with an average amplitude of 1.2~mmag; this really adds nothing to what was known from the work by \citet{p92}.  More interesting is the probable detection of an lpDNO.  A spike at 177.0~s shows above the red noise, this signal is at $5.483 \times P_{DNO}$, and is probably an lpDNO. 

\subsection{X Leo}
\label{sec:xleo}
X Leo is a U Gem star with $P_{orb}=3.946$~h \citep{xleoper}, a quiescent $V$ magnitude of $\sim$16.5, and outbursts of amplitude $\sim$4~mag roughly every 20~d \citep{xleotype}.  \citet{p87} report a quasi-coherent oscillation at 160~s in an outburst light curve of X Leo, and note that the oscillation is red compared to the disc. 

\begin{figure*}
 \includegraphics[width=140mm]{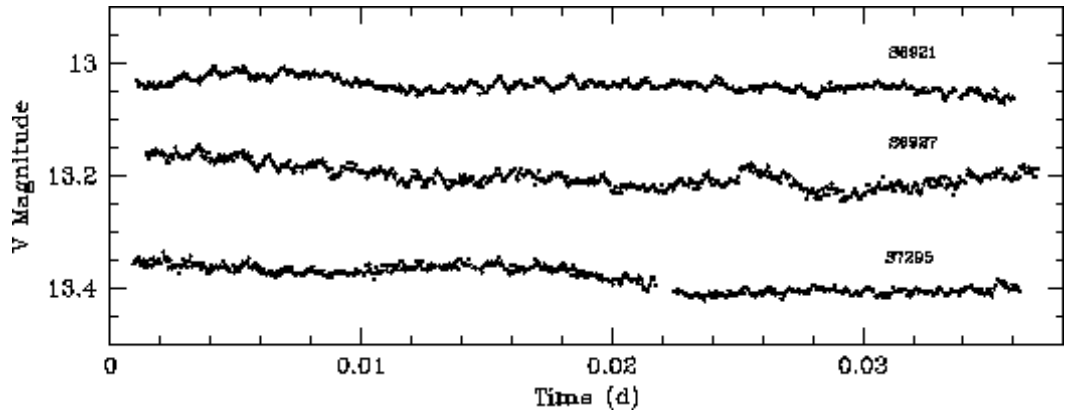}
 \caption {A sample of X Leo light curves in which some of the larger amplitude lpDNO cycles can easily be seen.  The S6921 and S7295 data are displaced vertically by +0.1 and +0.3~mag respectively, and all three light curves are arbitrarily shifted along the horizontal axis.}
 \label{fig:xleolp}
\end{figure*}

\begin{figure*}
 \includegraphics[width=140mm]{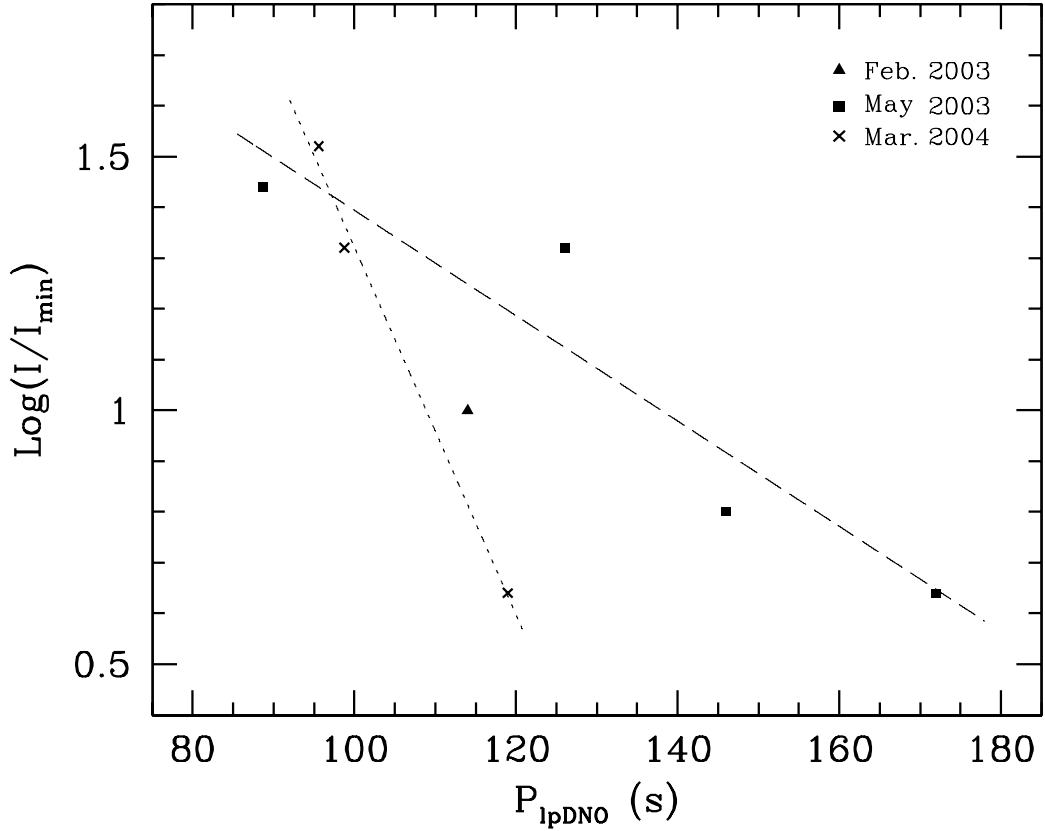}
 \caption{$\log(I/I_{min})$ as a function of lpDNO period for the three different outbursts of X Leo.  The dashed line is a least squares fit to the four points (squares) of the May 2003 outburst; the dotted line is a fit to the March 2004 data (crosses).  A single point (triangle) represents the February 2003 outburst.}
 \label{fig:xleopl}
\end{figure*}

We found that X Leo has commonly occurring lpDNOs with periods of the order of 100~s; these can in all cases be seen directly in the light curves; Fig. \ref{fig:xleolp} shows three examples.  X Leo is the first CV to show, for an lpDNO, a clear period-luminosity relation; all other observations of lpDNOs seem to indicate that $P_{lpDNO}$ has either no, or only very weak dependence on $\dot{M}$.  This relation is plotted in Fig. \ref{fig:xleopl} for the May 2003, and March 2004 outbursts, with the single point for February 2003 added.  In other systems, lpDNOs have been observed to double in frequency, but that is clearly not the case here.  The dashed and dotted lines in Fig \ref{fig:xleopl} are least squares fits to the May 2003 and March 2004 period-luminosity functions respectively.  
These give $s\equiv d\log(P_{lpDNO}/\mathrm{s}) / d\log(I/I_{min})=-0.29$ for the May 2003 outburst, and $s=-0.11$ for the March 2004 outburst.  Despite the factor of almost 3 difference between the gradients of the period-luminosity relation in the two outbursts, $s$ in both cases falls within the range measured for DNOs (see e.g. \citealt{bible}).

\begin{figure*}
 \includegraphics[width=140mm]{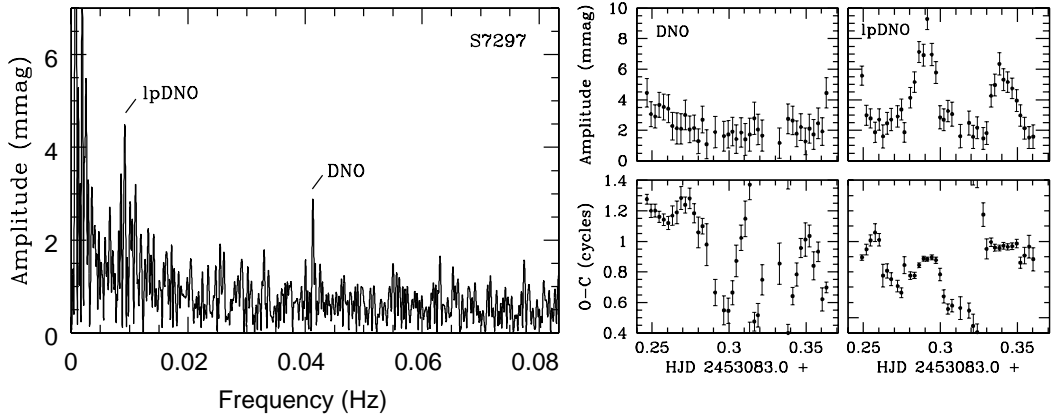}
 \caption {Left: The Fourier transform of an hour long section of the 19 March 2004 light curve of X Leo (run S7297), with DNO and lpDNO features marked.  Right: Amplitude and $O-C$ phase diagrams for the DNOs and lpDNOs in run S7297.  The DNO phase is calculated relative to a fixed period of 24.21~s, and that of the lpDNO relative to a fixed period of 118.5~s.}
 \label{fig:s7297ftomc}
\end{figure*}

In addition to the lpDNOs, the light curves taken on 14 and 19 March 2004 contained DNOs.  In the first of these, the DNO was present at a detectable (but very low) amplitude for only about 35~min.; the period was 12.9~s.  The 19 March 2004 light curve, on the other hand, is one of the clearest examples of a simultaneously present DNO and lpDNO.  A Fourier transform of part of this run is shown as the left hand panel of Fig. \ref{fig:s7297ftomc}.  The amplitude and $O-C$ diagrams for the oscillations (right hand side of Fig. \ref{fig:s7297ftomc}) show no correlation between the amplitude and phase variability of the DNO and lpDNO. 

Run S6932 contains a modulation on a preferred time scale of around 25~min., and of fairly large amplitude.  This kind of variability was only observed once, and it is therefore not clear whether it is a QPO.

\section{Oscillations in quiescent systems}

\subsection{V893 Sco}
\label{sec:v893sco}
Only very few dwarf novae below the period gap are not well credentialed SU UMa stars, V893 Sco is the best studied of these\footnote{V893 Sco is an ecliptic object, and hence unobservable for about two months every year, so that superoutbursts may have been overlooked.} (see e.g. \citealt{v893type}).  It is an eclipsing dwarf nova with an orbital period of 1.823~h \citep{v893per}, and outbursts every $\sim$30~d which takes it from a photographic magnitude of 14.5 to 10.6 \citep{v893type}.

V893 Sco has been observed to have QPOs at around 350~s (Paper III; \citealt{v893per}), and DNOs at 25.2~s (Paper III).

\begin{figure*}
 \includegraphics[width=140mm]{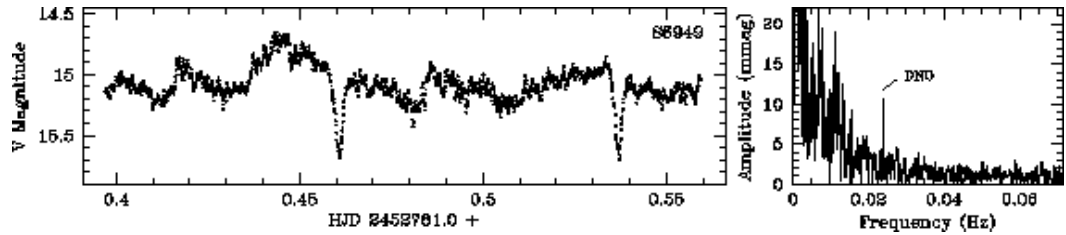}
 \caption {Left: The light curve of V893 Sco on 21 May 2003 (run S6949), showing the system at quiescence.  Right: The Fourier transform of a 0.9~h section of the light curve (starting just after the first eclipse).  A DNO with a period of 41.76~s is indicated.}
 \label{fig:v893lcft}
\end{figure*}

We observed this system in May 2003 on the decline from outburst, and into quiescence; surprisingly, it was again above minimum only two days after reaching quiescence (see Table \ref{tab:obs} for details).  No rapid oscillations were found in any of the outburst light curves, but run S6949, taken at minimum, contains a DNO with a period of 41.76~s---a rare example of a DNO in quiescence.  Fig. \ref{fig:v893lcft} shows the relevant light curve, together with a Fourier transform of a shorter section of the light curve.  The DNO was present for only 0.9~h, and the signal-to-noise is too low to enable a more thorough investigation of its behaviour. 

\subsection{WX Hyi}
WX Hyi is an SU UMa type dwarf nova, its normal outbursts have average maxima of $V=12.7$~mag and recur roughly every 14~d; the average $V$ magnitude at supermaximum is 11.4, and the supercycle is about 195~d.  The system's visual magnitude varies in the range 15.0--14.2 at quiescence \citep{bate86}.  \citet{s+v81} spectroscopically measured the orbital period to be 1.7955~h.  QPOs at $\sim$1140~s and, on a different occasion,  $\sim$1560~s were observed during the decline from outburst by \citet{p33}; we have previously detected a 185~s QPO present at the same time as a DNO with a period of 19.4~s, while WX Hyi was at $V=14.5$ (Paper III).

\begin{figure*}
 \includegraphics[width=140mm]{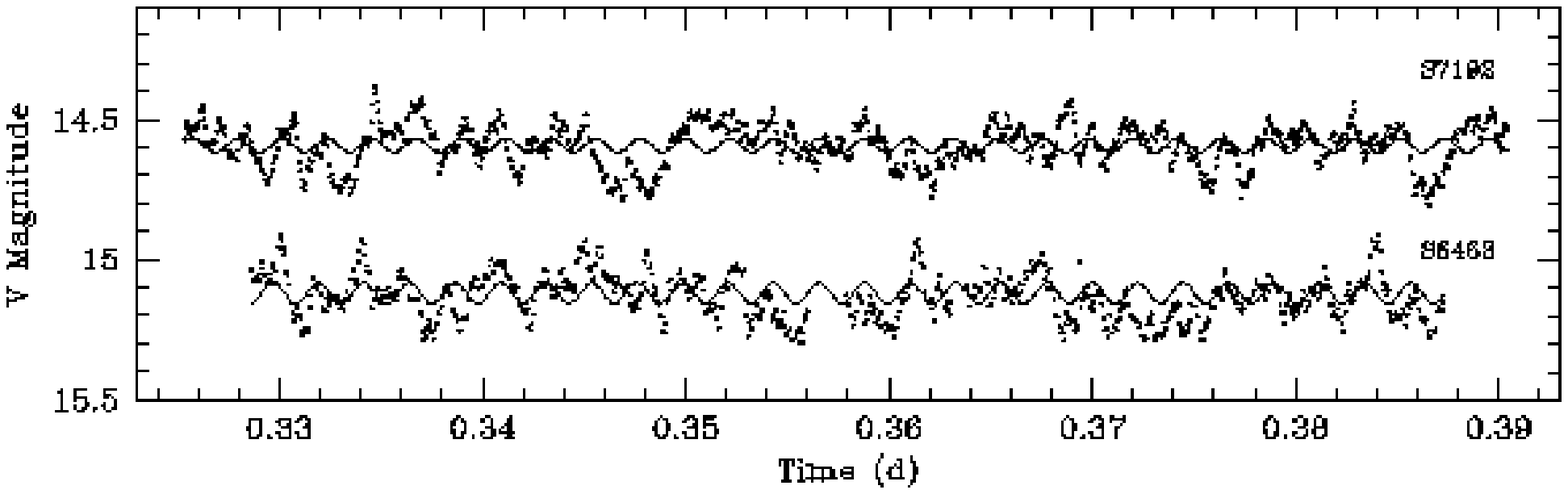}
 \caption{Sections of two WX Hyi light curves; slow variations were removed from both, and run S6463 is vertically shifted by $+0.2$~mag.  The best-fit sine functions to the QPOs (for these sections) are over-plotted, the QPO periods are 193.5~s and 190.2~s for run S6463, and S7192 respectively.}
 \label{fig:wxhyiqpo}
\end{figure*}

Table \ref{tab:obs} lists two observations of WX Hyi in quiescence.  Neither light curve contained DNOs, but QPOs where detected in both; they are displayed in Fig. \ref{fig:wxhyiqpo}.  This is only the second example of a QPO at quiescence.  On both occasions the QPO period was near 190~s, which is similar to the 185~s QPO reported in Paper III.

Note that for the three $\sim$100~s QPOs the period luminosity relation has the same sense as that for DNOs, although this may not mean much, since they were not observed during the same outburst cycle.  The QPOs illustrated in Fig. \ref{fig:wxhyiqpo}, and also the WX Hyi QPO shown in Paper III, keep remarkably good phase.

\section{Discussion and Summary}
\label{sec:discuss}

We have detected rapid, quasi-coherent oscillations in six CVs for the first time, and have observed new aspects of the phenomenology of oscillations in seven more systems.

The QPOs in V1193 Ori show signs of an underlying periodic signal, and are similar to kilosecond QPOs seen in V442 Oph, RX J1643.7+3402, and several more SW Sex stars, which have been interpreted as a low coherence manifestation of the underlying stable rotation of a magnetic white dwarf \citep{p94}.  However, V1193 Ori shares none of the other characteristics of SW Sex stars. 

RU Peg displays QPO periods ranging smoothly over almost an order of magnitude; this may call into question the existence of two clearly separate sets of QPOs discussed by W04.  In HX Peg, however, a QPO and kilosecond QPO have been detected simultaneously, implying that these are distinct phenomena.

Many contradictory and confusing claims about periodic modulations in V426 Oph have been made.  The detection of a DNO in this system casts further doubt on its status as an intermediate polar.

An lpDNO in X Leo was found to follow a period-luminosity relation similar to what is observed for DNOs.  Very little is presently known about lpDNOs; there are probably many features of their phenomenology that have not yet surfaced.  As soon as lpDNOs were identified they were found to be fairly common---although not as common as DNOs---and have up until the present been seen in about 24 CVs (in around one half of these systems the oscillations were at first reported as QPOs or DNOs).  It is perhaps surprising that lpDNOs remained unrecognised for so long; however, there is nowhere in the earlier literature an amplitude spectrum that makes the existence of a third class of oscillations as clearly evident as e.g.\ the Fourier transform shown as fig. 3 of W04.

Detections of DNOs, QPOs, and lpDNOs in dwarf novae at minimum light are review in W04 (see also \citealt{brianst}); to these can now be added a DNO in V893 Sco in quiescence, and QPOs in WX Hyi at minimum.  Rapid oscillations are probably more commonly present in quiescent dwarf novae than is currently thought; it is certainly true that oscillations occur much more frequently in outburst than in quiescence, which means that they are not often looked for in quiescent light curves.  For this reason, the observational record is at this point so heavily biased towards outbursting dwarf novae that it is impossible to make a secure estimate of the frequency with which oscillations occur in quiescence.

In summary, we have presented
\begin{enumerate}
\item the first detection of rapid, quasi-coherent oscillations in WW Cet, BP CrA, BR Lup, HP Nor, AG Hya, and V1193 Ori,
\item the detection of additional classes of oscillations in V426 Oph, V1159 Ori, and X Leo,
\item large evolution of a QPO period over 10 nights of the decline from outburst in RU Peg,
\item observations of normal QPOs and kilosecond QPOs simultaneously present in HX Peg,
\item a period-luminosity relation for lpDNOs in X Leo, and
\item the detection of oscillations in quiescence in the dwarf novae WX Hyi and V893 Sco.
\end{enumerate}

\section*{Acknowledgements}
MLP acknowledges financial support from the National Research Foundation and the Department of Labour.  BW's research is funded by the University of Cape Town.  PAW is supported by the National Research Foundation and by the University of Cape Town.


\label{lastpage}

\clearpage

\clearpage

\begin{thebibliography}{99}
\bibitem[\protect\citeauthoryear{Bateson}{1982a}]{cra1}Bateson F.M., 1982a, Publ. Var. Star Sect. R. Astron. Soc. New Zealand, 10, 12
\bibitem[\protect\citeauthoryear{Bateson}{1982b}]{cra2}Bateson F.M., 1982b, Publ. Var. Star Sect. R. Astron. Soc. New Zealand, 10, 24
\bibitem[\protect\citeauthoryear{Bateson \& McIntosh}{1986}]{bate86}Bateson F.M., McIntosh R., 1986, Publ. Var. Star Sect. R. Astron. Soc. New Zealand, 14, 1
\bibitem[\protect\citeauthoryear{Bond et al.}{1987}]{v1193type}Bond H.E., Grauer A.D., Burnstein D., Marzke R.O., 1987, PASP, 99, 1097 
\bibitem[\protect\citeauthoryear{Bruch \& Engel}{1994}]{hprange}Bruch A., Engel A., 1994, A\&AS, 104, 7
\bibitem[\protect\citeauthoryear{Bruch, Steiner \& Gneiding}{Bruch et al.}{2000}]{v893per}Bruch A., Steiner J.E., Gneiding C.D., 2000, PASP, 112, 237
\bibitem[\protect\citeauthoryear{Echevarr\'{i}a, Costero \& Michel}{Echevarr\'{i}a et al.}{1993}]{aghya}Echevarr\'{i}a J., Costero R., Michel R., 1993, A\&A, 275, 201
\bibitem[\protect\citeauthoryear{Hamuy \& Maza}{1986}]{v1193}Hamuy M., Maza J., 1986, IBVS, 2867
\bibitem[\protect\citeauthoryear{Hellier et al.}{1990}]{hell90}Hellier C., O'Donoghue D., Buckley D., Norton A., 1990, MNRAS, 242, 32P
\bibitem[\protect\citeauthoryear{Hessman}{1988}]{hess88}Hessman F.V., 1988, ApJS, 72, 515
\bibitem[\protect\citeauthoryear{Kato, Matsumoto \& Uemura}{Kato et al.}{2002}]{v893type}Kato T., Matsumoto K., Uemura M., 2002, IBVS, 5262
\bibitem[\protect\citeauthoryear{Homer et al.}{2004}]{v426}Homer L., Szkody P., Raymond J.C., Fried R.E., Hoard D.W., Hawley S.L., Wolfe M.A., Tramposch J.N., Yirak K.T., 2004, ApJ, 610, 991
\bibitem[\protect\citeauthoryear{Honeycutt et al.}{1998}]{hxpegtype}Honeycutt R.K., Robertson J.W., Turner G.W., Mattei J.A., 1998, PASP, 110, 676
\bibitem[\protect\citeauthoryear{Jablonski \& Cieslinski}{1992}]{jc92}Jablonski F.J., Cieslinski D., 1992, A\&A, 259, 198
\bibitem[\protect\citeauthoryear{Kuulkers et al.}{1991}]{p33}Kuulkers E., Hollander A., Oosterbroek T., van Paradijs J., 1991, A\&A, 242, 401
\bibitem[\protect\citeauthoryear{Marsh \& Horne}{1998}]{p19}Marsh T.R., Horne K., 1998, MNRAS, 299, 921
\bibitem[\protect\citeauthoryear{Mauche}{2002}]{p17}Mauche C.W., 2002, ApJ, 580, 423
\bibitem[\protect\citeauthoryear{Mennickent \& Sterken}{1998}]{lupper}Mennickent R.E., Sterken C., 1998, PASP, 110, 1032
\bibitem[\protect\citeauthoryear{Middleditch \& C\'{o}rdova}{1982}]{p87}Middleditch J., C\'{o}rdova F., 1982, ApJ, 255, 585
\bibitem[\protect\citeauthoryear{Nogami et al.}{1995}]{nkmh95}Nogami D., Kato T., Masuda S., Hirata R., 1995, IBVS, 4155
\bibitem[\protect\citeauthoryear{Norton \& Watson}{1989}]{n+w89}Norton A.J., Watson M.G., 1989, MNRAS, 237, 853
\bibitem[\protect\citeauthoryear{O'Donoghue}{1987}]{luptype}O'Donoghue D., 1987, Ap\&SS, 136, 247
\bibitem[\protect\citeauthoryear{O'Donoghue}{1995}]{uctccd}O'Donoghue D., 1995, Baltic Astron., 4, 519
\bibitem[\protect\citeauthoryear{Papadaki et al.}{2004}]{papad}Papadaki C., Boffin H.M.J., Cuypers J., Stanishev V., Kraicheva Z., Genkov V., 2004, in Hilditch R.W., Hensberge H., Pavlovski K., eds, ASP Conf. Ser. Vol. 318, Spectroscopically and Spatially Resolving the Components of Close Binary Stars. Astron. Soc. Pac., San Francisco, p. 399
\bibitem[\protect\citeauthoryear{Patterson, Robinson \& Nather}{Patterson et al.}{1977}]{p11}Patterson J., Robinson E.L., Nather R.E., 1977, ApJ, 214, 144
\bibitem[\protect\citeauthoryear{Patterson et al.}{1995}]{p92}Patterson J., Jablonski F., Koen C., O'Donoghue D., Skillman D.R., 1995, PASP, 107, 1183
\bibitem[\protect\citeauthoryear{Patterson et al.}{2002}]{p94}Patterson J., Fenton W.H., Thorstensen J.R., Harvey D.A., Skillman D.R., Fried R.E., Monard B., O'Donoghue D., Beshore E., Martin B., Niarchos P., Vanmunster T., Foote J., Bolt G., Rea R., Cook L.M., Butterworth N., Wood M., 2002, PASP, 114, 1364
\bibitem[\protect\citeauthoryear{Pretorius}{2004}]{thesis}Pretorius, M.L., 2004, MSc dissertation, University of Cape Town
\bibitem[\protect\citeauthoryear{Ringwald}{1994}]{hxpegper}Ringwald F.A., 1994, MNRAS, 270, 804
\bibitem[\protect\citeauthoryear{Ringwald, Thorstensen \& Hamwey}{Ringwald et al.}{1994}]{v1193per}Ringwald F.A., Thorstensen J.R., Hamwey R.M., 1994, MNRAS, 271, 323
\bibitem[\protect\citeauthoryear{Ringwald et al.}{1996}]{wwcetper}Ringwald F.A., Thorstensen J.R., Honeycutt R. K., Smith R.C., 1996, AJ, 111, 2077
\bibitem[\protect\citeauthoryear{Robertson, Honeycutt \& Turner}{Robertson et al.}{1995}]{rht95}Robertson J.W., Honeycutt R.K., Turner G.W., 1995, PASP, 107, 443
\bibitem[\protect\citeauthoryear{Rosen et al.}{1994}]{r94}Rosen S.R., Clayton K.L., Osborne J.P., McGale P.A., 1994, MNRAS, 269, 913
\bibitem[\protect\citeauthoryear{Schoembs \& Vogt}{1981}]{s+v81}Schoembs R., Vogt N., 1981, A\&A, 97, 185
\bibitem[\protect\citeauthoryear{Shafter \& Harkness}{1986}]{xleoper}Shafter A.W., Harkness R.P., 1986, AJ, 92, 658
\bibitem[\protect\citeauthoryear{Stover}{1981}]{rupegper}Stover R.J., 1981, ApJ, 249, 673
\bibitem[\protect\citeauthoryear{Szkody}{1986}]{szk86}Szkody P., 1986, ApJ, 301, L29 
\bibitem[\protect\citeauthoryear{Szkody}{1987}]{szk87}Szkody P., 1987, ApJS, 63, 685 
\bibitem[\protect\citeauthoryear{Szkody \& Mattei}{1984}]{xleotype}Szkody P., Mattei J.A., 1984, PASP, 96, 988 
\bibitem[\protect\citeauthoryear{Szkody, Kii \& Osaki}{Szkody et al.}{1990}]{szk90}Szkody P., Kii T., Osaki Y., 1990, AJ, 100, 546
\bibitem[\protect\citeauthoryear{Thorstensen et al.}{1997}]{ttbr97}Thorstensen J.R., Taylor C.J., Becker C.M., Remillard R.A., 1997, PASP, 109, 477
\bibitem[\protect\citeauthoryear{van Teeseling}{1997}]{p30}van Teeseling A., 1997, A\&A, 324, L73
\bibitem[\protect\citeauthoryear{Vogt \& Bateson}{1982}]{hptype}Vogt N., Bateson F.M., 1982, A\&AS, 48, 383
\bibitem[\protect\citeauthoryear{Warner}{1987c}]{wwcetty}Warner B., 1987c, MNRAS, 227, 23
\bibitem[\protect\citeauthoryear{Warner}{1995}]{bible}Warner B., 1995, Cataclysmic Variable Stars. Cambridge Univ. Press, Cambridge
\bibitem[\protect\citeauthoryear{Warner}{2004}]{rev}Warner B., 2004, PASP, 116, 115 (W04)
\bibitem[\protect\citeauthoryear{Warner \& Nather}{1988}]{v1193phot} Warner B., Nather R.E., 1988, IBVS, 3140
\bibitem[\protect\citeauthoryear{Warner \& Robinson}{1972}]{p38}Warner B., Robinson E.L., 1972a, Nature Phys. Sci., 239, 2
\bibitem[\protect\citeauthoryear{Warner \& Woudt}{2002}]{dno2} Warner B., Woudt P.A., 2002, MNRAS, 335, 84 (Paper II)
\harvarditem{Warner \& Woudt}{2004}{brianst}Warner B., Woudt P.A., 2005, in Hamuery J.-M., Lasota J.-P., eds, The Astrophysics of Cataclysmic Variables and Related Objects. ASP Conf. Ser. Vol. 330, Astron. Soc. Pac., San Francisco, p. 227
\bibitem[\protect\citeauthoryear{Warner, O'Donoghue \& Wargau}{Warner et al.}{1989}]{typsa}Warner B., O'Donoghue D., Wargau W., 1989, MNRAS, 238, 73
\bibitem[\protect\citeauthoryear{Warner, Woudt \& Pretorius}{Warner et al.}{2003}]{dno3}Warner B., Woudt P.A., Pretorius M.L., 2003, MNRAS, 344, 1193 (Paper III)
\bibitem[\protect\citeauthoryear{Woudt \& Warner}{2002}]{dno1}Woudt P.A., Warner B., 2002, MNRAS, 333, 411 (Paper I)

\end{thebibliography}
\end{document}